%% file: main.tex
\documentclass[aps,twocolumn,10pt,prc,floatfix,showpacs,preprintnumbers,amsmath,amssymb,nofootinbib,superscriptaddress,]{revtex4-1}
\input{packages}

\begin{document}

\input{commands}
\title{Nuclear energy density functionals grounded in ab initio calculations}
\input{authors}

\input{abstract}

\maketitle 

\input{introduction}
\input{ab_initio}

\input{method}

\input{results}

\input{conclusion}
\input{acknowledgement}

\appendix
\input{appendix}

\bibliography{bibliography.bib} 

\end{document}

%% file: packages.tex
\usepackage[normalem]{ulem}
\usepackage{wasysym}
\usepackage{color}
\usepackage{graphicx}
\usepackage{subfig}
\usepackage{dcolumn}    
\usepackage{multirow, booktabs }
\usepackage{comment}

\usepackage[format=plain,labelsep=period]{caption}
\usepackage{physics, mathtools}   
\usepackage [ english ]{ babel }

\usepackage{bigstrut}
\usepackage{amsmath,amsfonts,amsthm,bm}
\usepackage{CJK}
\usepackage[pdfstartview=FitH,
            CJKbookmarks=true,
            bookmarksnumbered=true,
            bookmarksopen=true,
            colorlinks,
            pdfborder=001,
            linkcolor=blue,
            anchorcolor=blue,
            citecolor=blue,
            urlcolor=blue,
            ]{hyperref}

%% file: commands.tex
\newcommand{\varR}{\mathbf{r}}
\newcommand{\varX}{\mathbf{x}}

\newcommand{\Sch}{ Schr\"{o}dinger }
\newcommand{\coeffSch}{-\frac{\hbar^2}{2m}}

\newcommand{\etal}{\textit{et al.} }
\newcommand{\redchi}{\chi^2_{red}}

%% file: authors.tex
\author{F. Marino}
\email{francesco.marino@unimi.it}
\affiliation{Dipartimento di Fisica ``Aldo Pontremoli'', Universit\`a degli Studi di Milano, 20133 Milano, Italy}
\affiliation{INFN,  Sezione di Milano, 20133 Milano, Italy}

\author{C. Barbieri}
\affiliation{Dipartimento di Fisica ``Aldo Pontremoli'', Universit\`a degli Studi di Milano, 20133 Milano, Italy}
\affiliation{INFN,  Sezione di Milano, 20133 Milano, Italy}

\author{G. Colò}
\affiliation{Dipartimento di Fisica ``Aldo Pontremoli'', Universit\`a degli Studi di Milano, 20133 Milano, Italy}
\affiliation{INFN,  Sezione di Milano, 20133 Milano, Italy}

\author{A. Lovato}
\affiliation{ Physics Division, Argonne National Laboratory, Argonne, Illinois 60439, USA }
\affiliation{INFN-TIFPA Trento Institute of Fundamental Physics and Applications, 38123 Trento, Italy}

\author{F. Pederiva}
\affiliation{Dipartimento di Fisica, University of Trento, via Sommarive 14, I–38123, Povo, Trento,
Italy}
\affiliation{INFN-TIFPA Trento Institute of Fundamental Physics and Applications, 38123 Trento, Italy}

\author{X. Roca-Maza}
\affiliation{Dipartimento di Fisica ``Aldo Pontremoli'', Universit\`a degli Studi di Milano, 20133 Milano, Italy}
\affiliation{INFN,  Sezione di Milano, 20133 Milano, Italy}

\author{E. Vigezzi}
\affiliation{INFN,  Sezione di Milano, 20133 Milano, Italy}

%% file: abstract.tex
\begin{abstract}
We discuss the construction of a nuclear Energy Density Functional (EDF) from ab initio calculations, and we advocate the need of a
methodical 
approach that is free from {\em ad hoc} assumptions. The equations of state (EoS) of symmetric nuclear and pure neutron matter are computed using the chiral NNLO$_{\rm sat}$ and the phenomenological AV4$^\prime$+UIX$_{c}$ Hamiltonians as inputs in the Self-consistent Green's Function (SCGF) and Auxiliary Field Diffusion Monte Carlo (AFDMC) methods, respectively. We propose a convenient parametrization of the EoS as a function of the Fermi momentum and fit it on the SCGF and AFDMC calculations. We apply the ab initio-based EDF to carry out an analysis of the binding energies and charge radii of different nuclei in the local density approximation. The NNLO$_{\rm sat}$-based EDF produces encouraging results, whereas the AV4$^\prime$+UIX$_{c}$-based one is farther from experiment. 
Possible explanations of these different behaviors are suggested, and the importance of gradient and spin-orbit terms is analyzed. Our work paves the way for a practical and systematic way to merge ab initio nuclear theory and DFT, while at the same time it sheds light on some of the critical aspects of this procedure. 
\end{abstract}

%% file: introduction.tex
\section{Introduction}
\label{sec: introduction}
The need to tackle the very complex nuclear many-body problem has inspired 
dramatic advances in the so-called ab initio methods in recent years~\cite{LEIDEMANN2013,computational_nuclear,Hergert_2020}. These approaches
aim at solving the many-nucleon Schr\"{o}dinger equation in an exact or systematically improvable way by using a realistic model for the nuclear interaction in the vacuum. Examples of these approaches are the Green's Function Monte Carlo (GFMC) and Auxiliary Field Diffusion Monte Carlo (AFDMC) \cite{carlson2015,Lynn2019,gandolfi2020}, Self-consistent Green's function (SCGF) \cite{Dickhoff2004,Barbieri2017,Rios2020,soma2020}, Coupled-cluster \cite{Hagen_2014,Hagen:2013yba,computational_nuclear}, In-Medium Similarity Renormalization Group \cite{Hergert_2020,Hergert2016} and Many-body perturbation theory \cite{Drischler2019,tichai2020}.
Successful nuclear structure calculations have been performed for low- and medium-mass nuclei \cite{LEIDEMANN2013,carlson2015,soma2020_chiral,Hergert_2020}, as well as in infinite nuclear matter \cite{Rios2020,Lonardoni2020,carbone2020} and neutron stars \cite{neutron_stars,Gandolfi2015}. 
Although ab initio theory can now approach masses of $A\approx 140$ \cite{Arthuis_2020}, its predictive power is affected by the large computational cost and full-scale studies of heavy nuclei are still out of reach.

In the heavy-mass region of the nuclear chart, the method of choice is Density Functional Theory (DFT). 
Originally introduced in condensed matter, DFT is a hugely popular method that finds application in several areas of physics, ranging from quantum chemistry \cite{ParrYang1994,martin_2004,becke2014,burke2012} to nuclear physics \cite{colo2020,Schunck2019,bender2003,Nakatsukasa2016,Dobaczewski2010,Stone_2007}. In this latter case, it represents the only approach that allows to cover almost the whole nuclear chart \cite{Dobaczewski2010,Schunck2019,colo2020}, with the partial exception of very light nuclei, and to study both the ground state (g.s.) and, in its time-dependent version, the excited states \cite{Nakatsukasa2016}.
In principle, DFT provides an exact formulation of the many-body problem based on the Hohenberg-Kohn theorems \cite{Hohenberg1964,ParrYang1994,Dobaczewski2010}, which state that all observables, starting from the total energy, can be expressed in a unique way as a functional of the one-body density (or of spin-densities and other generalized densities \cite{Chomaz}). No hints are given, however, on the actual form of such functional, which is dubbed as energy density functional (EDF).
Hence, in practice, DFT turns out to be an approximate, albeit very powerful, method. 
For the nuclear case in particular,
most relativistic \cite{vretenar2005} and non-relativistic \cite{bender2003,Schunck2019,colo2020} functionals
are designed in an empirical manner. A reasonable ansatz for the functional form is chosen
and its actual parameters are fitted on experimental observables such as radii and masses of finite nuclei, or pseudo-observables such as the saturation density of symmetric nuclear matter \cite{Chabanat1997_part1,Schunck2019}. The available EDFs are overall successful \cite{colo2020,Dobaczewski2010}, e.g. the experimental binding energies are reproduced on average within 1-2 MeV and charge radii within 0.01-0.02 fm. 
However, it is unclear how to further improve the performance of traditional EDFs \cite{unedf}. 
Despite attempts to frame DFT as an effective field theory (EFT), we still lack guiding principles for the systematic improvement of nuclear EDFs \cite{Furnstahl_2020}. 
Existing EDFs are affected by uncontrolled extrapolation errors when
applied to systems for which scarce data are available, like neutron-rich nuclei or superheavy nuclei \cite{Grasso2019,colo2020}.
To solve these issues, a rethinking of the strategy to build the EDFs is in order.

The success of electronic DFT stems from the fact that the 
Coulomb force is known, and 
the dominant contribution to the total energy is the simple Hartree term, while the exchange-correlation energy
is normally a relatively small correction \cite{martin_2004,ParrYang1994}.
In nuclear physics the situation is much more involved since the nuclear interaction, unlike the Coulomb force, has a rather complicated operator structure and is not uniquely known \cite{Machleidt2020}. 
Nevertheless, the  current state of ab initio calculations for nuclear systems makes it timely to ground nuclear DFT into the underlying theory, as it has already been done for Coulomb systems.
While there are already some attempts along this line
\cite{Furnstahl_2020,Dobaczewski_2016,Salvioni_2020}, e.g. using the Density Matrix Expansion (DME) \cite{navarro_perez2018,zurek2020}, or constraining specific terms of the EDFs on ab initio calculations \cite{baldo2008,Baldo2013,rocamaza2011,gambacurta2011,Papakonstantinou2018,Bulgac2018}, in this work we propose a novel and more systematic approach.
This approach is well-established in electronic DFT, and known as the ``reductionist'', or non-empirical, research program \cite{perdew,burke2012,perdew2017,perdew1996}. 
It aims at constructing a ``Jacob's ladder'' \cite{perdew} of increasingly more sophisticated EDFs, where one relies as much as possible on exact properties and ab initio calculations and as little as possible on fits to the empirical data. 
In the first rung of the ladder, the Local Density Approximation (LDA), an EDF that depends on the number densities alone is derived from the Equations of State (EoS) of uniform matter, i.e.
the homogeneous electron gas (HEG) in the electronic case \cite{ParrYang1994,martin_2004}, and symmetric nuclear matter (SNM) plus pure neutron matter (PNM) in the nuclear case. 
The second rung introduces surface terms within the so-called Gradient Approximation (GA), possibly constrained by ab initio simulations of non-uniform systems and/or by general principles. 

 In nuclear theory, the reductionist strategy is essentially unexplored. 
The purpose of this work is to make a first systematic study of the LDA step for nuclear physics. 
The nuclear Hamiltonian is not uniquely known (Sec. \ref{sec: nuclear interactions}) and, at variance with the HEG case, the nuclear matter EoS is still not firmly established.
Therefore, 
different microscopic interactions with the respective EoS, and possibly different many-body methods, should be explored.
In this work, we consider two distinct Hamiltonians. In AFDMC calculations we employ the simplified phenomenological AV4$^\prime$+UIX$_{c}$ \cite{lonardoni2017} potential, as it combines a reasonable accuracy with a relatively modest computational effort that allows us to compute nuclei as large as $^{90}$Zr. Second, we use the established and accurate chiral interaction NNLO$_{\rm sat}$ from Ref.~\cite{Ekstrom2015} as input for SCGF computations \cite{carbone2020}. We argue that the SNM and PNM EoS 
obtained either from AV4$^\prime$+UIX$_{c}$ or from NNLO$_{\rm sat}$ are well parametrized in powers of the Fermi momentum $k_F$ and we further perform a model selection procedure. 
The LDA EDFs resulting from the best parametrizations are used for g.s. calculations of closed-subshell nuclei and compared to ab initio results and to the experiment.
Moreover, a very preliminary exploration of the GA level is outlined. 

This paper is structured as follows. Sec. \ref{sec: intro ab initio} provides an introduction to ab initio theory, presenting Hamiltonians (Sec. \ref{sec: nuclear interactions}) and many-body methods (SCGF in Sec. \ref{sec: intro scgf}, AFDMC in \ref{sec: intro afdmc}). 
Sec. \ref{sec: method} is devoted to the construction of the nuclear EDFs.
The general framework is outlined in Sec. \ref{sec: intro nuclear edfs}. The parametrization of the ab initio EoS is examined in Sec. \ref{sec: param ab initio eos}. Then, LDA EDFs and GA EDFs are discussed in Sec. \ref{sec: edf lda}.
In Sec. \ref{sec: results}, the results are presented. The ab initio EoS determined with NNLO$_{\rm sat}$ and AV4$^\prime$+UIX$_{c}$ are shown and interpolated (Sec. \ref{sec: results infinite matter}). 
The corresponding LDA EDFs and GA EDFs are applied to finite nuclei in Sec. \ref{sec: results finite nuclei} and \ref{sec: results ga edf}, respectively.
Lastly, concluding remarks are presented in Sec. \ref{sec: conclusion}. 


%% file: ab_initio.tex
\section{Ab initio}
\label{sec: intro ab initio}

\subsection{Nuclear interactions}
\label{sec: nuclear interactions}
An effective description of nuclear systems at low energies can assume the nucleons as degrees of freedom and model their interactions through a non-relativistic Hamiltonian, which includes two- (NN) and three-nucleon (3N) (and possibly many-nucleon) potential terms:
\begin{equation}
    H = T + \sum_{i<j} V_{ij}^{NN} + \sum_{i<j<k} V_{ijk}^{3N} + ...
\end{equation}
Determining the nuclear interaction is still a partially open problem, but realistic models of the nuclear force, that are fitted to reproduce accurately two- or few-body observables, e.g. NN scattering phase shift or the binding energy of the deuteron, do exist.
Interactions are constructed in a phenomenological way \cite{Wiringa1995,carlson2015}, or by making use of chiral EFTs~\cite{Machleidt2020,Epelbaum2009,Machleidt_2016,epelbaum2020}. 
Chiral forces are derived in an order-by-order, low-momentum expansion consistent with the QCD symmetries.
They are defined in momentum space, although coordinate-space versions of the so-called local forces have been put forward~\cite{Gezerlis:2013ipa, Lynn:2014zia,Piarulli:2016vel, Piarulli:2017dwd,Lonardoni2018}. Since they are naturally cut off at high momenta by regulators \cite{rios2017}, they elude the problem of handling the hard core, i.e. the strongly repulsive short-range behaviour of the phenomenological potentials \cite{soma2020}.
The calculations performed in this work make use of the chiral NNLO$_{\rm sat}$ and the phenomenological AV4$^\prime$+UIX$_{c}$ interactions.  

NNLO$_{\rm sat}$ \cite{Ekstrom2015} is a chiral force that has been found to give a good simultaneous reproduction of binding energies and radii, as well as densities \cite{Arthuis_2020}, up to medium-mass nuclei, while it also predicts a saturation point close to the empirical region. In spite of some drawbacks, e.g. the symmetry energy around and above saturation is underestimated \cite{carbone2020}, NNLO$_{\rm sat}$ is among the best performing model chiral Hamiltonians.

Argonne interactions are widely employed phenomenological potentials \cite{Wiringa1995,Wiringa2002,carlson2015}. The most sophisticated of them is the Argonne $v_{18}$ (AV18) NN force, which contains 18 spin/isospin operators. Simplified versions, more amenable to many-body calculations, have been devised~\cite{Wiringa:2002ja}. Denoted as  AVn$^\prime$, these interactions contain a subset of $n$ operators and are refitted in order to reproduce as many two-nucleon properties as possible. 
Together with the NN interaction, a three-nucleon force (3NF) has to be introduced to reproduce the spectrum of light nuclei and saturation properties of infinite nucleonic matter \cite{sammarruca2020}. 
In Refs.~\cite{lonardoni2017,lonardoni2013hypernuclei} it was found that the simple AV4$^\prime$, which comprises only four operators:
\begin{equation}
    O_{ij}^{p=1,..,4} =
    \left[ 1, \mathbf{\sigma}_i \cdot \mathbf{\sigma}_j \right] \otimes 
    \left[ 1, \mathbf{\tau}_i \cdot \mathbf{\tau}_j \right],
\end{equation}
complemented with the central term of the Urbana IX 3N interaction (UIX$_c$), yields reasonable ground-state energies of light and medium-mass nuclei --- the binding energies deviate by about $10\%$ from experiment. AV4$^\prime$+UIX$_{c}$ is therefore interesting for this exploratory work, since it allows to carry out QMC studies of nuclear matter and larger nuclei than currently possible with fully-realistic phenomenological interactions. Moreover, the fact that AV4$^\prime$ does not contain tensor or spin-orbit operators greatly simplifies the solution of the many-body problem with the AFDMC method.

\subsection{SCGF}
\label{sec: intro scgf}
The Self-consistent Green's function (SCGF) method \cite{Dickhoff2004,Barbieri2017,soma2020,Rios2020} provides a non-perturbative and systematically improvable solution to the\Sch equation for a system of $A$ interacting fermions that is rooted on the concept of many-body propagators, also known as Green's functions.
The central quantity is the one-particle propagator
\begin{equation}
g_{\alpha \beta}(\omega) = -\frac{i}{\hbar}
 \int dt \, e^{i \omega t} \, \bra{\Psi_0} T[ c_\alpha(t) c^\dagger_\beta(0)]  \ket{\Psi_0} \; ,
\label{eq:def_gab}
\end{equation}
where greek letters label the states of a complete orthonormal single-particle basis, $c_\alpha(t)$~($\,c^\dagger_\alpha(t)\,$) are the creation (annihilation) operators of a particle in state $\alpha$ at time $t$, and $T[\ldots]$ is the time ordering operator. 
The propagator $g_{\alpha \beta}(\omega)$ provides access to all one-body observables and to the ground state energy; moreover, it gives information on the neighbouring $A \pm 1$ nuclei, for example by providing the one-nucleon addition and removal energies \cite{soma2020_chiral}. 
The interest in SCGF lies in the wealth of information that can be accessed in a single stage, as well as in its computational cost scaling polinomially with the number of particles \cite{Barbieri2017}. 
It is nonetheless a demanding method and fully converged results for g.s. energies are presently possible for mass numbers up to $A \approx 60-90$,  depending on the chosen Hamiltonian, and for nuclear matter~\cite{soma2020}. In a recent work \cite{Arthuis_2020}, though, charge radii and charge density distributions have been successfully obtained in a set of Sn and Xe isotopes up to $A=138$.

In SCGF, the dressed Green's function is determined by solving the Dyson equation, 
\begin{equation}
g_{\alpha \beta}(\omega) = 
g^{(0)}_{\alpha \beta}(\omega) + \sum_{\gamma \delta}
g_{\alpha \gamma}(\omega) \, \Sigma^*_{\gamma \delta}(\omega) \, g_{\delta \beta}(\omega) \, ,
\label{eq:Dyson}
\end{equation}
or the equivalent Gorkov equation when one deals with open shell nuclei \cite{soma2013}.
In Eq.~\eqref{eq:Dyson}, $g_{\alpha \beta}^{(0)}(\omega)$ is an unperturbed propagator. The  $\Sigma^*_{\alpha \beta}(\omega)$ is the irreducible self-energy which is suitably written in a form that does not depend on a choice of the reference state but only on $g_{\alpha \beta}(\omega)$ itself. The Dyson or Gorkov equations are in principle exact but the self-energy, $\Sigma^*_{\alpha \beta}(\omega)$, must be truncated in practical applications by keeping selected series of Feynman diagrams, hence resumming infinite subsets.
Nuclear matter is studied within the finite-temperature formalism, making use of the ladder approximation to the self-energy \cite{Carbone2014,Barbieri2017,Rios2020}. The limit $T=0$ is taken at the end of the calculation. 
State of the art computations in finite nuclei exploit the algebraic diagrammatic construction method at order $n$ (ADC(\emph{n})) \cite{Barbieri2017,soma2020}, which is based on enforcing the correct analytic properties of the self-energy. ADC(\emph{n}) is systematically improvable 
since, as the order \emph{n} is increased, the exact solution is approached and it would be recovered in the limit $n=+\infty$.
The best many-body truncations of the self-energy currently available  are ADC(3) in Dyson SCGF and ADC(2) in the Gorkov formalism \cite{soma2020}.

\subsection{AFDMC}
\label{sec: intro afdmc}
Diffusion Monte Carlo (DMC) \cite{Lynn2019,carlson2015,Foulkes2001} methods 
solve the many-body \Sch equation by using imaginary-time projection techniques to enhance the ground-state component $\ket{\Psi_0}$ of a starting trial wave function $\ket{\Psi_T}$
\begin{equation}
\label{eq: evolution dmc}
    \ket{\Psi_0} \propto \lim_{\tau \to + \infty } \ket{\Psi(\tau)} = \lim_{\tau \to + \infty } e^{-(H-E_T)\tau}  \ket{\Psi_T},
\end{equation}
where $\tau$ is the imaginary time and $E_T$ is an estimate of the ground state energy. 

The application of DMC to nuclear physics is complicated by the presence of spin and isospin operators in the Hamiltonian. The two variants of continuum DMC algorithms, the Green's function Monte Carlo (GFMC)~ \cite{carlson2015,gandolfi2020} and the auxiliary-field diffusion Monte Carlo (AFDMC)~\cite{Schmidt:1999lik,Lonardoni2018,gandolfi2020}, differ in the way they deal with the spin/isospin degrees of freedom. The GFMC method uses all of the $2^A\binom{A}{Z}$ spin-isospin components of the wave function and can treat highly-realistic phenomenological and chiral interactions~\cite{Lynn:2014zia,Lynn:2015jua,Piarulli:2016vel,Piarulli:2017dwd}, but it is currently limited to nuclei with up to $A=12$ nucleons. On the other hand, within the AFDMC, the spin-isospin degrees of freedom are described by single-particle spinors, the amplitudes of which are sampled using Monte Carlo techniques based on the Hubbard-Stratonovich transformation, reducing the computational cost from exponential to polynomial in $A$. However, some of the contributions characterizing fully-realistic nuclear forces, such as isospin-dependent spin-orbit contributions, cannot be treated in this way, yet. Hence, the AFDMC is limited to somewhat simplified interactions, but it can be applied to compute larger nuclei and nuclear matter. 

The starting point of AFDMC calculations is a trial wave function, which is commonly expressed as the product of a long-range component $\ket{\Phi}$ and of two- plus three-body correlations
\begin{equation}
\label{eq: scalar corr psi_t}
    \ket{\Psi_T} = \prod_{i<j} f^c_{ij} \prod_{i<j<k} f^c_{ijk}  \ket{\Phi}
\end{equation}
In the above equation, we assumed the correlations to be spin-isospin independent. This simplified ansatz, consistent with Refs.~\cite{Contessi:2017rww, lonardoni2017,Schiavilla:2021dun}, is justified by the fact that the AV4$^\prime$+UIX$_{c}$ Hamiltonian does not contain tensor or spin-orbit terms. 

In finite nuclei, $\ket{\Phi}$ is constructed by coupling different Slater determinants of single-particle orbitals in the $\ket{nlj m_j}$ basis so as to reproduce the total angular momentum, total isospin, and parity of the nuclear state of interest~\cite{gandolfi2020}. On the other hand, infinite nuclear matter is modeled by simulating a finite number of nucleons on which periodic-box boundary conditions (PBC) are imposed \cite{Sarsa:2003zu}. In this case, the single-particle states are plane waves with quantized wave numbers
\begin{equation}
    \mathbf{k} = \frac{2\pi}{L} \left( n_x, n_y, n_z \right) \qquad n_i = 0, \pm 1, \pm 2...
\end{equation}
where $L$ is the size of the box and the shell closure condition must be met in order to satisfy translational invariance. As a consequence, the number of nucleons in a box must be equal to the momentum space ``magic numbers'' (1, 7, 19, 27, 33, $\dots$) times the number of spin/isospin states: 2 for PNM, 4 for SNM. The equations of state of nuclear matter discussed in Sec.~\ref{sec: results infinite matter} are computed with 66 neutrons (PNM) and 76 nucleons (SNM) in a periodic box.

The AFDMC method has no difficulty in dealing with ``stiff" forces that can generate wave functions with high-momentum components. This is in contrast with remarkably successful many-body approaches that rely on a basis expansion~\cite{Barrett:2013nh,Jurgenson:2013yya,Hagen:2013yba,Hagen:2013nca}, which need relatively ``soft" forces to obtain converged calculations. However, like standard diffusion Monte Carlo algorithms, the AFDMC suffers from the fermion sign problem, which results in large statistical errors that grow exponentially with $\tau$. To control it, we employ the constrained-path (CP) approximation, as described in Refs.~\cite{Lonardoni2018,piarulli2020,gandolfi2020}. This scheme is believed to be accurate for Hamiltonians that do not include tensor or spin-orbit operators, as is the case for the AV4$^\prime$+UIX$_{c}$ potential. 

Expectation values of operators $\hat{O}$ that  do not commute with the Hamiltonian are evaluated by means of the mixed estimator \cite{carlson2015}
\begin{equation}   \expval{\hat{O}(\tau)} \approx 2 \frac{\mel{\Psi_T}{\hat{O}}{\Psi(\tau)}}{\braket{\Psi_T}{\Psi(\tau)} } -  \frac{\mel{\Psi_T}{\hat{O}}{\Psi_T}}{\braket{\Psi_T}{\Psi_T} }. \end{equation}
Also, charge radii are estimated from the proton radii with the formula $r_{ch}^2 = r_p^2 + (0.8\, \rm{fm})^2$.

%% file: method.tex
\section{Method}
\label{sec: method}

\subsection{Nuclear EDFs}
\label{sec: intro nuclear edfs}
The general structure of a non-relativistic nuclear EDF is described in depth in Refs. \cite{bender2003,Schunck2019,ring}.
In this section, the discussion is limited to even-even nuclei and to quasi-local EDFs, i.e. functionals that can be expressed as the volume integral of an energy density $\mathcal{E}(\varR)$ which is function of the local densities \cite{bender2003} and their gradients. Non-local EDFs such as Gogny ones are not treated. 
Moreover, for simplicity pairing terms are neglected.
Applications shall be limited to magic nuclei and to some closed-subshell ones. 

 Under these assumptions, the total energy is a functional of the time-even proton and neutron densities (number density $\rho_q(\varR)$, kinetic  density $\tau_q(\varR)$ and spin-orbit density $\mathbf{J}_q(\varR)$, with $q=n,p$) \cite{bender2003,Chabanat1997_part1} and reads:
\begin{equation}
\label{eq: edf basic structure}
    E = \int d\varR\, \mathcal{E}(\varR) = E_{kin} + E_{pot} + E_{Coul}
\end{equation}
The kinetic energy term is given by \cite{Chabanat1997_part1}
\begin{equation}
      E_{kin} = \int d\varR\, \mathcal{E}_{kin}(\varR) = \int d\varR\, \frac{\hbar^2}{2m} \tau_0(\varR) 
\end{equation}
The Coulomb contribution $E_{Coul}$ is treated in the standard local Slater approximation \cite{colo_skyrme}. 
The most general form of the potential term:
\begin{equation}
        E_{pot} = \int d\varR\, \mathcal{E}_{pot}(\varR)
\end{equation}
is reported in Eqs. (48-49) of Ref. \cite{bender2003},  and will be outlined in the next section. Neutron and proton densities have been recoupled into the isoscalar (t=0) and isovector (t=1) channels: isoscalar densities are total densities (e.g. $\rho_0=\rho_n+\rho_p$), while isovector densities account for proton-neutron differences ($\rho_1=\rho_n-\rho_p$).  
The coefficients of the various terms are all, in principle, functions of the density, although in practice most of them are set to a constant value \cite{Schunck2019}.
The mean field equations are then derived by relating the densities to the single-particle orbitals $\phi_j(\varR)$ and applying the variational principle \cite{ring}:
\begin{eqnarray}
\label{eq: skyrme HF eqs}
    \bigg[ 
    - \nabla \cdot \frac{\hbar^2}{2m^*_q(\varR)} \nabla + U_q(\varR) + U_{Coul}(\varR) \delta_{q,p} + \\
    \mathbf{W}_q(\varR) \cdot 
    \left( -i\right) \left( \nabla \cross \mathbf{\sigma} \right)
    \bigg] 
    \phi_j(\varR) = \epsilon_j \phi_j(\varR)
\end{eqnarray}
where
\begin{equation}
\label{eq: def mwan field}
    U_q = \fdv{E}{\rho_q} \qquad 
    \frac{\hbar^2}{2m_q^*(\varR)} = \fdv{E}{\tau_q}
    \qquad \mathbf{W}_q = \fdv{E}{\mathbf{J}_q}
\end{equation}
and $m_q^{*}(\varR)$, $U_q(\varR)$ and $\mathbf{W}_q(\varR)$ are called effective mass, mean field and spin-orbit potential, respectively.

\subsection{Parametrization of the ab initio EoS}
\label{sec: param ab initio eos}
Nuclear matter is an infinite and uniform system of nucleons interacting through the strong interaction only \cite{Gandolfi2015,Margueron2018,rocamaza2018}.
Nuclear matter is characterized by the nuclear Equation of State (EoS), which at zero temperature and for spin-unpolarized matter corresponds to the energy per nucleon $e=E/A$ as a function of the number densities of neutrons and protons, or of the total density $\rho$ and the isospin asymmetry parameter $\beta=\frac{\rho_n-\rho_p}{\rho}$.

 Uniformity implies that the orbitals are plane waves, leading to important simplifications of Eq. \eqref{eq: edf basic structure}.
In fact, the number and kinetic densities are uniform and related by \cite{ring}
\begin{equation}
\label{eq: tau_q in infinite matter}
    \tau_q = \frac{3}{5} k_{F,q}^2 \, \rho_q = \frac{3}{5} \left( 3\pi^2 \right)^{2/3} \rho_q^{5/3}.
\end{equation}
Moreover, the spin-orbit densities $\mathbf{J}_q$, as well the derivatives of the density ($\nabla \rho_q=\Delta \rho_q =0$), vanish.
Thus only the $\rho$ and $\rho \tau$ terms contribute to the nuclear matter energy, while the gradient ($\rho\Delta\rho$), spin-orbit ($\rho \nabla \cdot \mathbf{J}$) and tensor ($\mathbf{J}^2$) terms are non-vanishing in non-uniform systems only, such as nuclei, neutron drops or semi-infinite matter.

 These considerations suggest the following regrouping of the potential density $\mathcal{E}_{pot}$ for a generic nuclear system
\begin{equation}
\label{eq: def pot term EDF}
    \mathcal{E}_{pot} = \mathcal{E}_{bulk} + \mathcal{E}_{surf},
\end{equation}
where
\begin{equation}
\mathcal{E}_{bulk} = \sum_{t=0,1} \, \big(  C_t^{\rho} \rho_t^2 +
C_t^{\tau}  \rho_t \tau_t \big)
\end{equation}
and 
\begin{equation}
\mathcal{E}_{surf} = \sum_{t=0,1} \, \big(  C_t^{\Delta \rho} \rho_t \Delta \rho_t + C_{t}^{J} \mathbf{J}_{t}^2 + C_t^{\nabla J} \rho_t \nabla \cdot \mathbf{J}_t  \big).
\end{equation}
These terms are called here bulk and surface contributions, respectively. Infinite matter probes only the bulk contributions, while surface terms are active in non-uniform systems. 

The problem of determining the nuclear EoS is still partially unsettled~\cite{Gandolfi2015,Tews2020,drischler2020,Rios2020}, at stark variance with the case of the homogeneous electron gas in condensed matter, whose EoS has been well-known for decades from Monte Carlo calculations, as well as from analytical results in the low- and high-density limits \cite{martin_2004}. 
By contrast, the choice of the nuclear Hamiltonian~\cite{carbone2020,Drischler2019,sammarruca2020, Lonardoni2020} and, to a lesser extent, of the many-body method \cite{piarulli2020}, still impacts our knowledge of the theoretical nuclear EoS.
Therefore, it is important to test our method on different EoS, in order to better understand its potentialities and limitations.

Before using it as input to an EDF, the EoS must be parametrized.
First of all, it is convenient to represent the energy per particle $e(\rho,\beta)$ as the sum of the kinetic energy per particle of the Fermi gas $t(\rho,\beta)$ and of a potential term $v(\rho,\beta)$ \cite{Margueron2018,Papakonstantinou2018}, consistently with the EDF structure \eqref{eq: edf basic structure}:
\begin{equation}
    e(\rho,\beta) = t(\rho,\beta) + v(\rho,\beta),
\end{equation}
where
\begin{equation}
\label{eq: kin en per particle}
    t(\rho,\beta)= \frac{t_{sat}}{2} \left[ 
    \left( 1+\beta\right)^{\frac{5}{3}} +\left( 1-\beta\right)^{\frac{5}{3}}
    \right] 
    \left( \frac{\rho}{\rho_{sat}}\right)^{\frac{2}{3}},
\end{equation} 
\begin{equation}
\label{eq: def t sat}
    t_{sat}= \frac{3\hbar^2}{10m} \left( \frac{3\pi^2}{2}\right)^{\frac{2}{3}} \rho_{sat}^{\frac{2}{3}} 
    \approx 22 \, \rm{MeV},
\end{equation}
and $t_{sat}$ is the Fermi gas kinetic energy per particle of SNM at saturation density, $t_{sat}=t(\rho=\rho_{sat},\beta=0)$.

Next, an ansatz for the expression of the potential energy per particle $v(\rho,\beta)$ as a function of both $\rho$ and $\beta$ must be chosen. Presently, ab initio methods are mostly applied to PNM ($\beta=1$) and SNM ($\beta=0$).
Hence, the dependence on $\beta$ must be extrapolated from the limiting cases $\beta=0$ and $\beta=1$. Due to the isospin invariance of the nuclear force, odd powers of $\beta$ vanish. Moreover, neglecting terms in $\beta^4$ is deemed accurate for densities close to saturation even for large asymmetries \cite{rocamaza2018}. This quadratic dependence is adopted here too.

As far as the $\rho$-dependence is concerned, one can reasonably expect that a limited number of powers of $\rho$ should suffice for reproducing the theoretical EoS, see e.g. Refs. \cite{Margueron2018,Lonardoni2020}. 
While a Taylor expansion in powers of the density is simple and useful \cite{Margueron2018,Baldo2013}, we argue that a better option is to postulate that the potential term be a polynomial of the Fermi momentum $k_F$, or equivalently of $\rho^{1/3}$ \cite{Papakonstantinou2018,Bulgac2018}. Heuristic motivations are the followings: from a practical perspective, it grants a greater flexibility than a $\rho$-expansion, to which it may eventually reduce as a special case. 
Also, it is known on an empirical basis that local EDFs need fractional powers of the density
in order to get satisfactory predictions of the nuclear incompressibility \cite{garg2018,bender2003}, thus using $k_F$ instead of $\rho$ as an expansion variable is also in keeping with this latter necessity. 
Lastly, if the EoS is thought as arising from a diagrammatic expansion, then powers of the Fermi momentum should appear naturally \cite{Papakonstantinou2018,kaiser2002}. 

Combining the above assumptions, one can then write
\begin{equation}
\label{eq: potential series}
    v(\rho,\beta) = \sum_{\gamma=1/3\ldots 6/3} \, c_\gamma(\beta) \rho^\gamma = \sum_{\gamma=1/3\ldots 6/3} \left[
    c_{\gamma,0} + c_{\gamma,1} \beta^2
    \right] \rho^\gamma
\end{equation}
where $c_{\gamma,0} \equiv c_\gamma(\beta=0)$ and $c_{\gamma,1} \equiv c_\gamma(\beta=1) - c_\gamma(\beta=0)$. 
Up to this point, the model is still quite general. The only condition is that $\gamma$'s have to be of the form integer/3.
Now, we do not choose the potential a priori \cite{Papakonstantinou2018}, but, in order to determine how many terms and which powers should enter the potential, we perform a model selection among all possible polynomials with at most six terms and $\gamma$ not larger than 6 (Eq. \eqref{eq: potential series}).
The following convention is employed: each model is identified by the exponents of the powers of $k_F$ or $\rho^{\frac{1}{3}}$ it contains. For example, we refer to the polynomial $c_{\frac{2}{3}}\rho^{\frac{2}{3}} + c_{\frac{5}{3}} \rho^{\frac{5}{3}} + c_2\rho^2$ by $(2,5,6)$.
Each model is fitted on the SNM and PNM data points and the optimal parameters are determined by minimizing
the mean squared error (MSE) ~\cite{chi_square,statistical-learning}
\begin{equation}
    \sigma^2 \left( c_{\gamma,0}, c_{\gamma,1}\right) =
    \frac{1}{N_{data}}
    \sum_{i=1}^{N_{data}} \Big[
    e(\rho_i,\beta_i)-e_i
    \Big]^2.
\end{equation}
Cross-validation is used to evaluate the out-of-sample error ~\cite{MEHTA20191,statistical-learning}, which we use to rank the different models. 
This is a more robust measure of goodness than the fit MSE or $\chi^2$ \cite{chi_square}.
The statistical analysis has been performed with the scikit-learn library ~\cite{scikit-learn}. 

\subsection{Construction of the ab initio-based EDFs}
\label{sec: edf lda}
The simplest way to define an EDF based on the infinite matter EoS is Local Density Approximation (LDA) \cite{Stone_2007,martin_2004,Baldo2013}. 
In LDA, one assumes that the same expression of the potential energy density valid in infinite matter holds for non-uniform densities $\rho_q(\varR)$ too.
This approximation is well-suited in particular for slowly varying density distributions, so that each small region of a generic (finite or infinite) system can be treated as a piece of bulk matter \cite{martin_2004}. LDA provides the following expression for the bulk energy density $\mathcal{E}_{bulk}(\varR)$:
\begin{equation}
    \mathcal{E}_{bulk}\left[ \rho(\varR), \beta(\varR) \right] =
    \rho(\varR)  v\left[ \rho(\varR), \beta(\varR) \right].
\end{equation}
The LDA EDFs read
\begin{equation}
    E_{LDA} = E_{kin} + E_{bulk} + E_{Coul}
\end{equation}
and Eq. \eqref{eq: skyrme HF eqs} simplifies, as $m^{*}=m$, $\mathbf{W}(\varR)=0$ and $U_q(\varR)=U_q^{bulk}(\varR)$, where
\begin{align}
    \label{eq: mean field LDA}
    & U_q^{bulk}(\varR) = \fdv{E_{bulk}}{\rho_q(\varR)} =
       \\ & 
     \sum_\gamma \Big[
    \left(  \gamma+1 \right) c_{\gamma,0}
    + \big(  \left( \gamma -1 \right) \beta(\varR) + 2\tau_z  \big) \beta(\varR) \, c_{\gamma,1}
    \Big] \rho^\gamma(\varR), \nonumber
\end{align}
for the potential term \eqref{eq: potential series} and $\tau_z=+1$ for neutron and $\tau_z=-1$ for protons. See App. \ref{app: derivation mean field} for the derivation.

While LDA is the main subject of this paper, eventually it is not sufficient to describe nuclear systems (Sec. \ref{sec: results finite nuclei}).
Moreover, we will show that the LDA EDFs based on either Hamiltonian give rather different outcomes. 
Thus, in order to better gauge LDA, we have decided to perform a preliminary analysis of a set of EDFs that include surface terms. 
These functionals, that we name gradient approximation (GA) EDFs, are made by complementing LDA with isoscalar and isovector density-gradient terms and a one-parameter spin-orbit contribution. 
It must be understood that these GA EDFs are treated at a very preliminary level. For instance, $\rho\tau$ terms, that are known to be important in nuclear DFT and produce an effective mass $m^* \neq m$, are not discussed. Also, no rigorous statistical analysis is performed and no attempt to derive the surface terms from ab initio is made. These important themes are left for future studies.

Our GA EDFs have the following form:
\begin{equation}
    E_{GA} = E_{LDA} + E_{surf}  
\end{equation}
where
\begin{align}
    E_{surf} = \int d\varR & \Bigg[ 
    \, \sum_{t=0,1} C_t^\Delta \rho_t \Delta \rho_t   \\
    & - \frac{W_0}{2} \left( \rho \nabla\cdot\mathbf{J} + \sum_q \rho_q \nabla\cdot\mathbf{J}_q \right) 
    \Bigg]  \, . \nonumber
\end{align}
Three parameters, $C_0^\Delta$, $C_1^\Delta$ and $W_0$, are introduced and are all assumed to be density-independent constants, as in widely used EDFs. 
The mean field equations \eqref{eq: skyrme HF eqs} hold, with $m^*=m$ and $U(\varR)=U_q^{bulk}(\varR)+U_q^{surf}(\varR)$, where
\begin{align}
    & \mathbf{W}_q(\varR) = \fdv{E_{surf}}{\mathbf{J}(\varR)} = \frac{W_0}{2} \left( \nabla \rho + \nabla \rho_q \right) 
    \\
    \label{eq: mean field surf}
    & U_q^{surf}(\varR) = \fdv{E_{surf}}{\rho_q} = \nonumber \\ 
    & = 2 C_0^\Delta \Delta\rho_0 + 2 C_1^\Delta \Delta\rho_1 \tau_z 
    - \frac{W_0}{2} \left( 
    \nabla \cdot \mathbf{J} + \nabla \cdot \mathbf{J}_q
    \right) 
\end{align}
and $U_q^{surf}$ is derived in App. \ref{app: mean field ga}. App. \ref{app: derivation rearrangement} is dedicated to the concept of rearrangement energy of the EDF. 

 To tune the surface terms, a grid search on the three parameter $C_0^\Delta$, $C_1^\Delta$ and $W_0$ is carried out, although full-fledged fits will be necessary in later works. 
To benchmark the quality of the EDF predictions, the root mean square (rms) errors of the binding energies and the charge radii for the GA EDFs
\begin{subequations}
\label{eq: rms errors}
\begin{align}
    \sigma_E(C_0^\Delta, C_1^\Delta, W_0) = \sqrt{ \frac{\sum_{k=1}^{n_E} \left( E_k^{th} - E_k^{exp} \right)^2 }{n_E}  }, \\ 
    \sigma_{r_{ch}}(C_0^\Delta, C_1^\Delta, W_0) = \sqrt{ \frac{\sum_{k=1}^{n_r} \left( r_k^{th} - r_k^{exp} \right)^2 }{n_r}  }.
\end{align}
\end{subequations}
are evaluated with respect to the experimental radii of $^{40}Ca$, $^{48}Ca$, $^{132}Sn$ and $^{208}Pb$ and the binding energies of $^{40}Ca$, $^{48}Ca$, $^{90}Zr$, $^{132}Sn$ and $^{208}Pb$ ~\cite{rocamaza2012}. 
All the DFT g.s. calculations are performed with the skyrme\_rpa code~\cite{colo_skyrme}, which has appropriately modified.

%% file: results.tex
\section{Results}
\label{sec: results}

\subsection{Nuclear matter fits}
\label{sec: results infinite matter}
\begin{figure*}[t]
    \begin{minipage}[t]{\columnwidth}
        \includegraphics[width=\columnwidth]{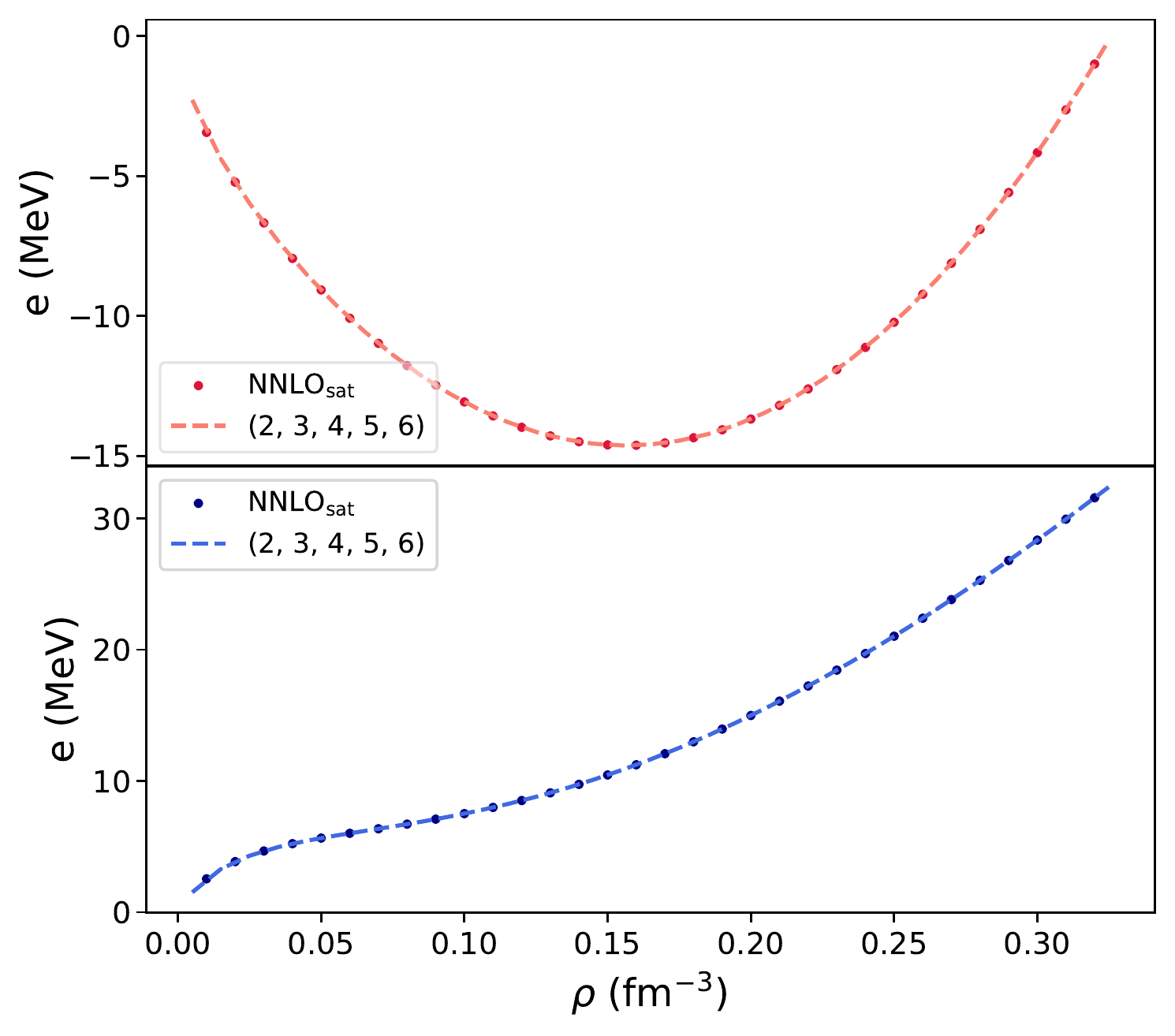}
    \caption{Dots: SNM and PNM EoS computed with the NNLO$_{\rm sat}$ interaction and the SCGF method. Dashed: model EoS $(2,3,4,5,6)$ (see text). }
    \label{fig: eos_nnlosat}
    \end{minipage}
    \begin{minipage}[t]{\columnwidth}
    \includegraphics[width=\columnwidth]{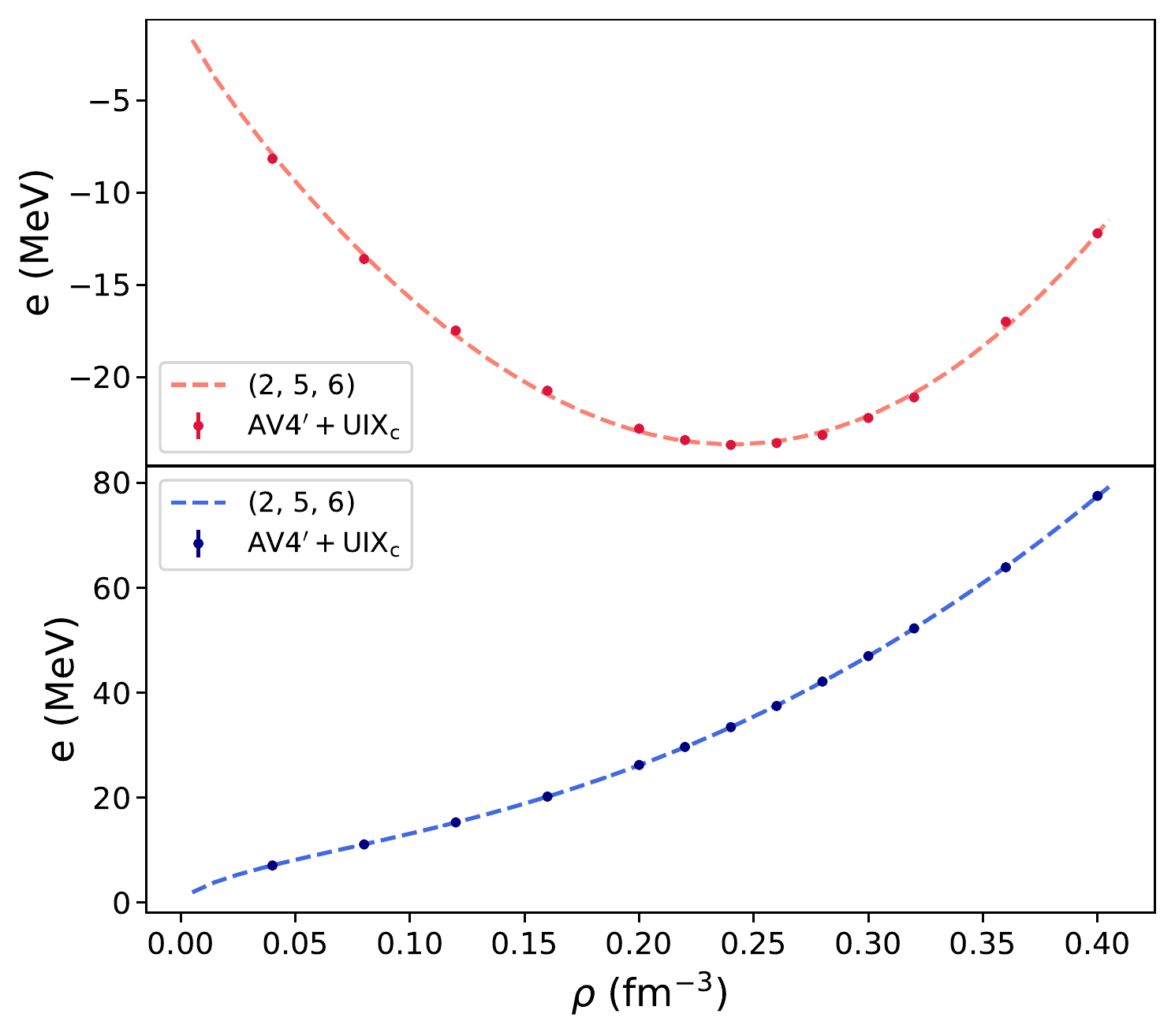}
    \caption{Dots: SNM and PNM EoS computed with the AV4$^\prime$+UIX$_{c}$ interaction and the AFDMC method. The AFDMC statistical error bars are shown. Dashed: model EoS $(2,5,6)$ (see text). }
   \label{fig: eos_av4p}
    \end{minipage}
\end{figure*}

The SCGF SNM and PNM equations of state have been computed for densities up to \mbox{$\rho=0.32$ fm$^{-3}$} with a step of \mbox{0.1 fm$^{-3}$} by means of the SCGF method. The SNM EoS saturates at \mbox{$\rho_{sat}$=0.15 fm$^{-3}$} and \mbox{E$_{sat}$=-14.7 MeV}. A 5-fold cross-validation has been performed and the average validation error has been used to select the best model (Sec. ~\ref{sec: intro nuclear edfs}). It turns out that the optimal choice is the model $(2,3,4,5,6)$, which achieves a very small \mbox{MSE=10$^{-8}$ MeV$^2$}. The NNLO$_{\rm sat}$ EoS and the $(2,3,4,5,6)$ model are shown in Fig. ~\ref{fig: eos_nnlosat}.

 The AV4$^\prime$+UIX$_{c}$ EoS has been calculated with the AFDMC method for several densities up to \mbox{0.40 fm$^{-3}$}.
To the best of our knowledge, this is the first application of AV4$^\prime$+UIX$_{c}$ to nuclear matter. The results are reported in Tab.~\ref{tab: eos av4p}.
The saturation point is located at an unusually high density (\mbox{$\rho$=0.24 fm$^{-3}$}) and low energy (\mbox{$E_{sat}$=-23.7 MeV}) and the 3N contribution is instrumental in allowing the SNM EoS to saturate; in fact, $\rm {AV4'}$ alone predicts no saturation before 0.50 fm$^{-3}$ \cite{baldo2012}. The smallest validation error (MSE=0.06 MeV$^2$) is achieved by the (2,5,6) model, which is shown in Fig.~\ref{fig: eos_av4p} together with the ab initio EoS.

 To sum up, parametrizing the nuclear EoS as a polynomial of the Fermi momentum has proved an effective ansatz. Two optimal models have been found, namely (2,3,4,5,6) for the NNLO$_{\rm sat}$ EoS and (2,5,6) for the AV4$^\prime$+UIX$_{c}$ EoS.

\begin{table}
\begin{ruledtabular}
\begin{tabular}{ 
    >{\centering}p{0.20\columnwidth} 
    >{\centering}p{0.20\columnwidth} 
    >{\centering\arraybackslash}p{0.20\columnwidth}
    }
     $\rho$ (fm$^{-3}$) & e (MeV) SNM & e (MeV) PNM \\
     \hline
     0.04 & -8.17 (1) & 7.062 (5) \\
     0.08 & -13.60 (1) & 11.075 (6) \\
     0.12 & -17.48 (1) & 15.278 (8) \\
     0.16 & -20.74 (2) & 20.20 (1) \\
     0.20 & -22.80 (1) & 26.23 (1) \\
     0.22 & -23.42 (2) & 29.66 (2) \\
     0.24 & -23.68 (3) & 33.44 (3) \\
     0.26 & -23.58 (3) & 37.47 (2) \\
     0.28 & -23.15 (3) & 42.12 (3) \\
     0.32 & -21.10 (3) & 52.26 (5) \\
     0.36 & -17.0 (1) & 63.91 (6) \\
     0.40 & -12.21 (8) & 77.51 (7)\\
\end{tabular}
\end{ruledtabular}
\caption{Energy per particle $e$ and standard errors (in parenthesis) computed with AFDMC and the AV4$^\prime$+UIX$_{c}$ interaction at several densities $\rho$ in both SNM and PNM. }
\label{tab: eos av4p}
\end{table}

\subsection{Predictions of the LDA EDFs in finite nuclei}
\label{sec: results finite nuclei}
Two LDA EDFs are derived from the (2,3,4,5,6) and (2,5,6) parametrizations of the  NNLO$_{\rm sat}$- and the AV4$^\prime$+UIX$_{c}$-based EoS (Sec. \ref{sec: results infinite matter}). Then, they are applied to closed-subshell nuclei and compared to experimental values, taken from Ref. \cite{charge_radii,binding_energies}, and to ab initio results. Full ab initio calculations are available for a set of nuclei up to $^{54}$Ca for NNLO$_{\rm sat}$ and $^{90}$Zr for AV4$^\prime$+UIX$_{c}$. Moreover, the NNLO$_{\rm sat}$ densities for $^{90}$Zr are available. 

\begin{figure*}
\begin{minipage}[t]{\columnwidth}
    \includegraphics[width=\columnwidth]{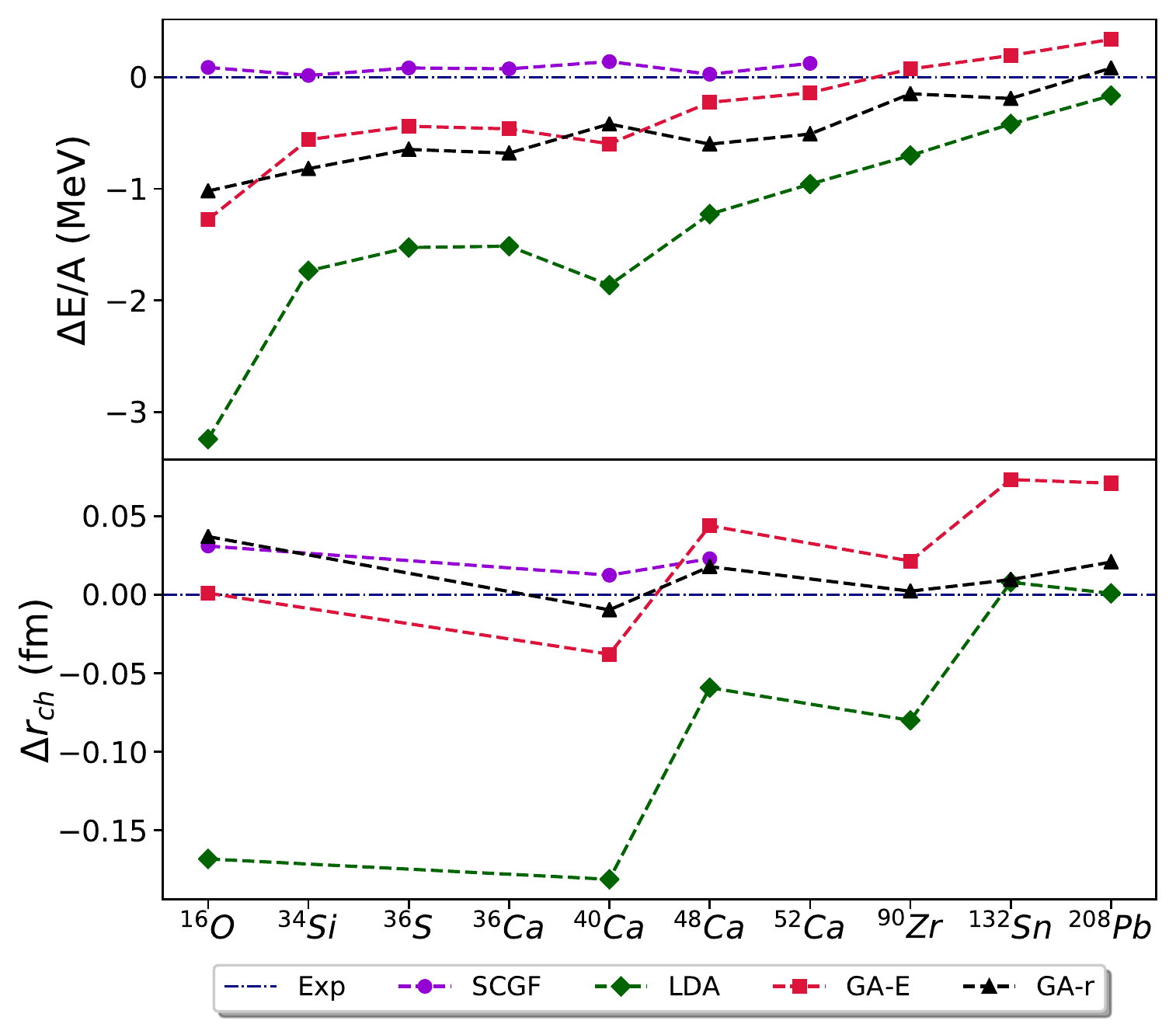}
    \caption{Discrepancy between the predicted energies per nucleon (top) and charge radii (bottom) and the corresponding experimental values for a set of closed subshell nuclei. Results obtained with the NNLO$_{\rm sat}$ interaction and with the LDA, GA-E and GA-r EDFs  are shown. The LDA EDF is derived from the (2,3,4,5,6) model EoS. The GA-E and GA-r EDFs are described in Sec.~\ref{sec: results ga edf}. }
    \label{fig: nnlosat energies and radii}
\end{minipage}
\hfill
\begin{minipage}[t]{\columnwidth}
    \includegraphics[width=\columnwidth]{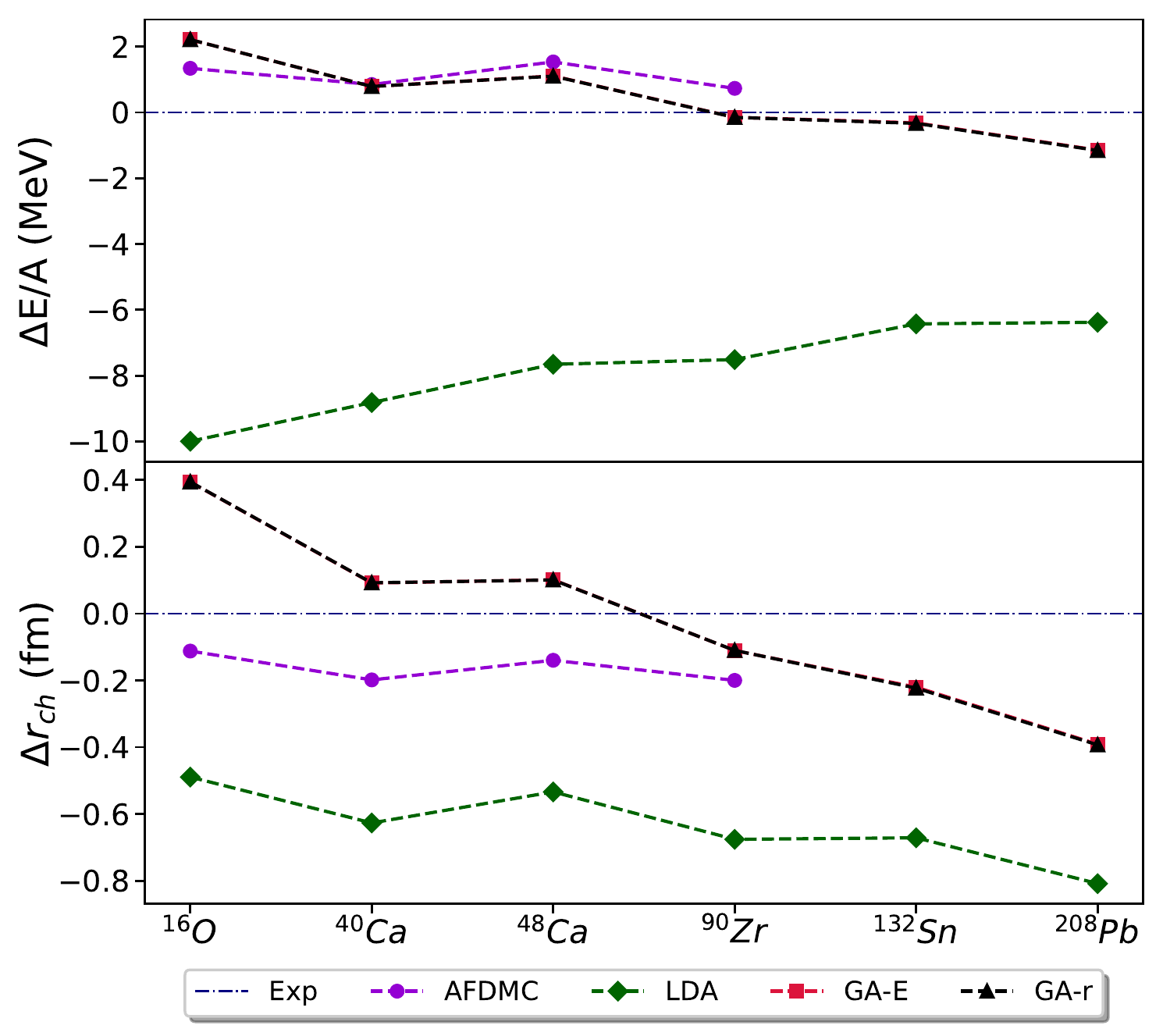}
    \caption{Same as  Fig. \ref{fig: nnlosat energies and radii} (note the different scale), but for results obtained with the AV4$^\prime$+UIX$_{c}$ interaction and with the LDA, GA-E and GA-r EDFs. The LDA EDF is based on the (2,5,6) EoS.   }
    \label{fig: av4p energies and radii}
\end{minipage}
\end{figure*}

The discrepancy between theory and experiment for energies per nucleon (top) and charge radii (bottom) are shown in Fig.~\ref{fig: nnlosat energies and radii} for  NNLO$_{\rm sat}$ and the (2,3,4,5,6) EDF, as well as the GA-E and GA-r EDFs introduced later on (Sec.~\ref{sec: results ga edf}).
On the one hand, we can appreciate that NNLO$_{\rm sat}$ predictions are very close to experiment. On the other hand, the LDA EDF, although less precise, exhibits interesting trends, since it enables to reproduce heavier nuclei, especially from $^{90}$Zr on, in a realistic way, with deviations smaller than 1 MeV/nucleon and 0.05 fm for the energies and radii, respectively. This is quite remarkable, as the LDA EDF incorporates only information on uniform matter. Also, it is unsurprising that light systems are less amenable to a local density treatment, since surface effects are known to play a larger role at small $A$'s.

 In Fig.\ref{fig: av4p energies and radii}, the deviation of the AV4$^\prime$+UIX$_{c}$, (2,5,6) EDF and GA-E and GA-r EDFs (Sec. \ref{sec: results ga edf}) predictions from experiment are shown.
The outcome is puzzling, since, while the ab initio results are overall decent, the LDA EDF (2,5,6) strongly overbinds all the nuclei considered, by $\approx$ 10 MeV per nucleon.
In addition, radii are underestimated with respect to both experiment and ab initio. Thus, in the case of the phenomenological interaction AV4$^\prime$+UIX$_{c}$, LDA alone has difficulties to capture the properties of the microscopic potential. 

\begin{figure*}[t]
    \includegraphics[height=0.60\textwidth,width=0.95\textwidth]{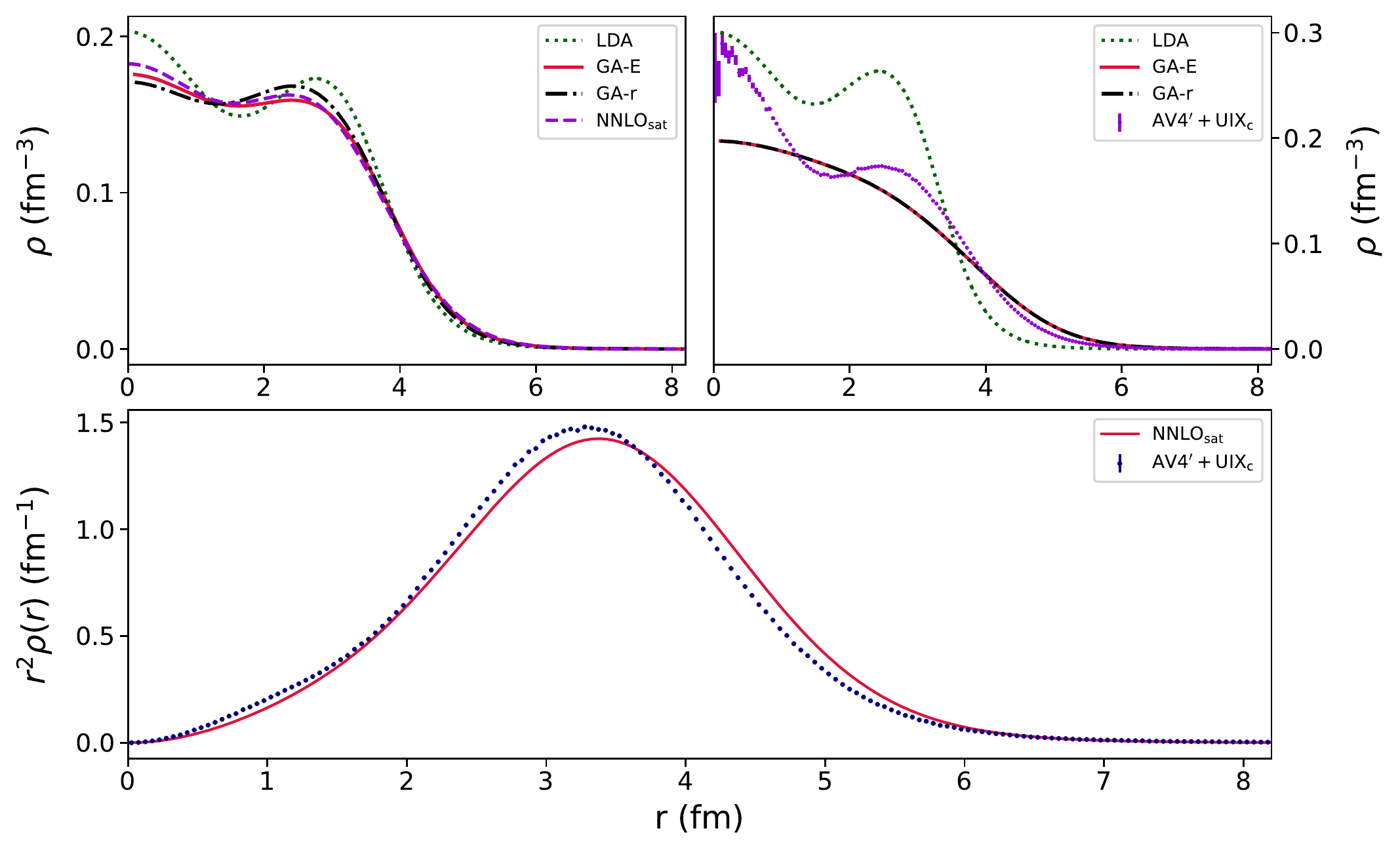}
    \caption{ 
    Ab initio and EDF (LDA, GA-E and GA-r) number densities, $\rho(r)$, for $^{48}$Ca computed using the NNLO$_{\rm sat}$ (top left) and AV4$^\prime$+UIX$_{c}$ (top right) Hamiltonians. See text for details. Note that for the AV4$^\prime$+UIX$_{c}$ case the GA-E and GA-r curves overlap closely.
    Bottom: ab initio number densities times the squared radius, $r^2 \rho(r)$,  obtained with NNLO$_{\rm sat}$ (full line) and AV4$^\prime$+UIX$_{c}$ (dotted).  }
    \label{fig: many_rho_ca48}
\end{figure*}

\begin{figure*}[t]
    \includegraphics[height=0.60\textwidth,width=0.90\textwidth]{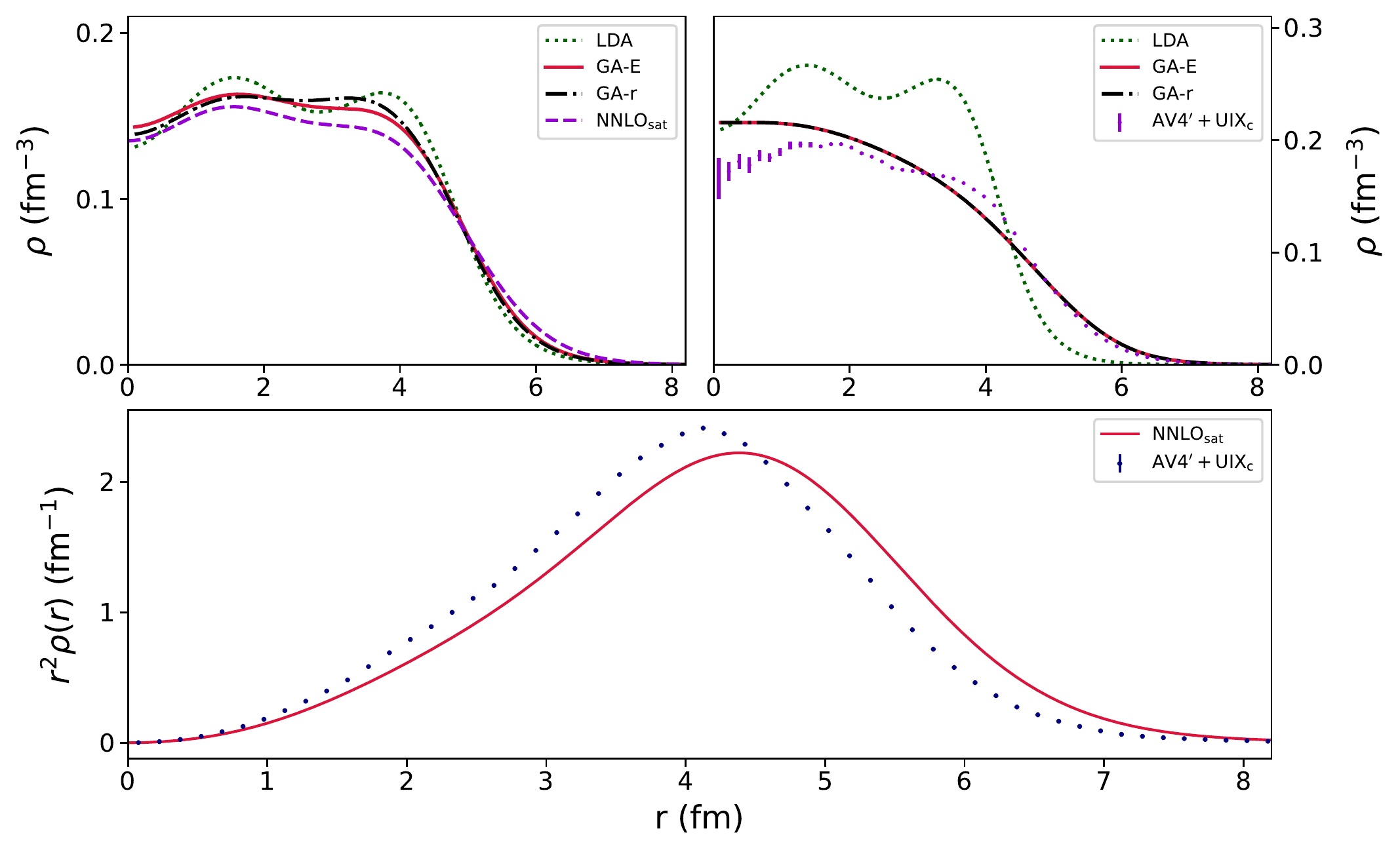}
    \caption{Ab initio and EDF (LDA, GA-E and GA-r) number densities of $^{90}$Zr obtained from NNLO$_{\rm sat}$ (top left) and AV4$^\prime$+UIX$_{c}$ (top right). 
    Note that for the AV4$^\prime$+UIX$_{c}$ case the GA-E and GA-r curves overlap closely.
    Bottom: ab initio results for $r^2 \rho(r)$  using NNLO$_{\rm sat}$ (full line) and AV4$^\prime$+UIX$_{c}$ (dotted).  }
    \label{fig: many_rho_zr90}
\end{figure*}

Number densities are then shown for the representative nuclei $^{48}$Ca (Fig. \ref{fig: many_rho_ca48}) and $^{90}$Zr (Fig. \ref{fig: many_rho_zr90}).
In the NNLO$_{\rm sat}$ case (top left), the (2,3,4,5,6) EDF density profile closely resembles the ab initio one, although it features slightly wider oscillation. 
In the AV4$^\prime$+UIX$_{c}$ case (top right), instead, the (2,5,6) EDF and ab initio number densities differ considerably, as LDA produces definitely steeper density profiles, consistently with predicting sensibly smaller radii. Also, it somewhat overestimates the central density.
In the bottom panel, the $^{48}$Ca ab initio densities weighted by the squared radius, $r^2 \rho(r)$ are compared.
The $r^2$ factor emphasizes that AV4$^\prime$+UIX$_{c}$ and NNLO$_{\rm sat}$ predict rather different density surfaces.
Roughly similar considerations hold for $^{90}$Zr, except that the discrepancy of AV4$^\prime$+UIX$_{c}$ with the (2,5,6) EDF, as well as with NNLO$_{\rm sat}$ is more accentuated. 

 In conclusion, the NNLO$_{\rm sat}$-based LDA EDF compares favourably with the experiment, in spite of its simplicity, and reproduces radii, energies and densities fairly well in magic nuclei, especially in the heavier ones. The AV4$^\prime$+UIX$_{c}$-based EDF, on the other hand, is less satisfactory and highlights even more clearly the necessity of introducing surface terms.

\subsection{Predictions of the GA EDFs }
\label{sec: results ga edf} 
In Sec. \ref{sec: edf lda}, simple GA EDFs have been introduced by complementing LDA with two gradient and one spin-orbit terms. In this section, the predictions of the GA EDFs based on NNLO$_{\rm sat}$ and AV4$^\prime$+UIX$_{c}$ are discussed. The parameters $C_0^\Delta$, $C_1^\Delta$ and $W_0$ are tuned by grid-searching over physically reasonable intervals and the results for the four EDFs that yield the smallest rms errors on binding energies or charge radii, called GA-E and GA-r for short, are shown. 
The three parameters are measured in \mbox{MeV fm$^{5}$}; from now on, for simplicity the dimension is omitted.

 In the case of the NNLO$_{\rm sat}$-based EDF (2,3,4,5,6), we have considered $C_0^\Delta$ and $C_1^\Delta$ in the intervals $[-40,0]$ and $[0,40]$ in steps of 5, while we have varied $W_0$ between 30 and 140 in steps of 10.
The smallest rms error on the energy is obtained for ($C_0^\Delta=-25$, $C_1^\Delta=10$, $W_0=50$), while charge radii are best reproduced for ($C_0^\Delta=-30$, $C_1^\Delta=25$, $W_0=140$) (Tab. \ref{tab: err ga}). 
The remarkable improvement over the LDA EDF can be appreciated by looking at energies and radii (Fig. \ref{fig: nnlosat energies and radii}).
In Fig. \ref{fig: many_rho_ca48} and \ref{fig: many_rho_zr90}, the effect of the gradient terms on the number densities is made clear by the disappearance of the oscillations which instead characterize the LDA densities. 
All considered, these GA EDFs are quite accurate, in spite of containing only three adjustable parameters, one of which ($C_1^\Delta$) is of minor importance for the g.s. properties. 
A full-fledged optimization would be necessary at a later stage of development to be truly competitive against the most sophisticated existing EDFs. 
However, the outcomes shown here are already a very encouraging starting point. 

 In the case of AV4$^\prime$+UIX$_{c}$-based EDFs, $C_0^\Delta$, $C_1^\Delta$ and $W_0$ have been varied in the intervals $[-200,-60]$, $[0,50]$ and $[0,150]$.
The smallest errors on radii and energies are obtained with ($C_0^\Delta=-155$, $C_1^\Delta=15$, $W_0=10$) (GA-r) and ($C_0^\Delta=-155$, $C_1^\Delta=0$, $W_0=10$) (GA-E), respectively. Highly repulsive gradient contributions are needed to compensate for the LDA overbinding excess. We also note that GA-r and GA-E have quite similar parameters and as a consequence lead to almost indistinguishable outcomes.
The GA EDFs perform significantly better than the LDA (2,5,6) EDF (Tab.~\ref{tab: err ga}).
Surface terms are effective in improving the binding energies, which are brought less than 1 MeV/A from experiment and quite close to AFDMC predictions (top panel of Fig.~\ref{fig: av4p energies and radii}). 
Note, however, that the scale is different from that of Fig.~\ref{fig: nnlosat energies and radii}, and that NNLO$_{\rm sat}$-based EDF are nonetheless more accurate.
Some problems persist, though, in particular concerning radii (bottom panel of Fig.~\ref{fig: av4p energies and radii}), which are still inaccurate for the nuclear DFT standards.

The different behaviour of the NNLO$_{\rm sat}$- and AV4$^\prime$+UIX$_{c}$-based EDFs calls for an explanation. The unrealistic saturation density of the AV4$^\prime$+UIX$_{c}$ EoS may correlate with LDA predictions being far from experiment. Also, it may explain, at least partially, why radii are not accurate, even at the GA level.
However, it cannot explain the large discrepancies of LDA (and GA, as far as radii are concerned) with respect to ab initio itself. 
We then suggest that the strong correlations induced by the hard core of the Argonne interaction may be difficult to catch within LDA, whereas the same scheme can be more successfully applied to the relatively soft NNLO$_{\rm sat}$ potential. The wide oscillations of the AV4$^\prime$+UIX$_{c}$ densities may witness the role of short-range correlations. 
Further investigations should focus on finding a quantitative measure of "hardness" appropriate for this problem \cite{rios2017}. 
\begin{table}
    \begin{ruledtabular}
    \begin{tabular}
    { p{0.18\columnwidth}
    >{\centering} p{0.2\columnwidth} | 
    >{\centering} p{0.21\columnwidth}
    >{\centering}p{0.18\columnwidth}
    >{\centering\arraybackslash}p{0.18\columnwidth} }
       & EDF & $\sigma_{E/A}$ (MeV) & $\sigma_{E}$ (MeV) & $\sigma_{r_{ch}}$ (fm) \\
       \hline
       \multirow{3}{*}{NNLO$_{\rm sat}$} & LDA & 1.07 & 59 & 0.10  \\
       & GA-E  & 0.30 & 13 & 0.03  \\
       & GA-r  & 0.34 & 21 & 0.01  \\ 
       \hline
       \multirow{3}{*}{AV4$^\prime$+UIX$_{c}$} & LDA & 7.4 &  800 & 0.67 \\ 
       & GA-E & 0.81 & 112 &  0.22 \\
       & GA-r & 0.77 & 120 &  0.21 \\ 
       \hline
    \end{tabular}
    \end{ruledtabular}
     \caption{Rms errors between theoretical predictions and experimental data on the binding energies energies per nucleon $\sigma_{E/A}$, total energies $\sigma_{E}$ and charge radii $\sigma_{r_{ch}}$. Calculations are performed with two sets of EDFs, based on the NNLO$_{\rm sat}$ and AV4$^\prime$+UIX$_{c}$ EoS. The LDA EDFs are based on the (2,3,4,5,6) EoS in the NNLO$_{\rm sat}$ case and on (2,5,6) EoS in the AV4$^\prime$+UIX$_{c}$ case.
     GA-E and GA-r stand for the GA EDFs that achieve the smallest discrepancy for binding energies and charge radii, respectively. The rms errors are evaluated on the same set of magic nuclei data as in Ref. \cite{rocamaza2012}.  }
    \label{tab: err ga}
\end{table}

%% file: conclusion.tex
\section{Conclusion}
\label{sec: conclusion}
In this work, we analyzed the first steps for defining nuclear EDFs that are directly based on ab initio theory. 
Specifically, we employed the Local Density Approximation (LDA) to derive EDFs from the ab initio nuclear EoS.
 The SNM and PNM EoS have been computed with two interactions and many-body methods, i.e. NNLO$_{\rm sat}$ with SCGF and AV4$^\prime$+UIX$_{c}$ with AFDMC. 
A parametrization of the EoS as a polynomial of the Fermi momentum has been proposed and shown to allow an adequate reproduction of the theoretical curves. The optimal models, 
then, have been used to define nuclear EDFs by means of LDA.
In order to better gauge LDA, we have also introduced a set of elementary GA EDFs, defined by complementing LDA EDFs with simple surface (density-gradient and spin-orbit) terms. 
Both LDA and GA EDFs have been tested on the g.s. of magic nuclei and compared to ab initio and experiment.

The NNLO$_{\rm sat}$-based EDFs show very interesting results. The simple LDA EDFs, that incorporate information on uniform matter only, fairly reproduce charge radii and binding energies. The agreement with the experiment is particularly good for the heavier nuclei.
Moreover, the predictive power improves considerably at the GA level, although we have attempted here only a preliminary GA implementation and further studies are required to compare with the most sophisticated empirical EDFs.
In contrast, the AV4$^\prime$+UIX$_{c}$ case is somewhat puzzling and the LDA appears to be much less satisfactory in our study. 
We observe that very repulsive gradient contributions are needed at the GA level and do make an important effect, significantly shrinking the discrepancy with AFDMC calculations. Still, some more issues persist; in particular, radii compare to experiment in an unsatisfactory way for the DFT standards. 

The different behaviour of NNLO$_{\rm sat}$- and AV4$^\prime$+UIX$_{c}$-based EDFs needs to be understood.
We suggest that LDA may struggle to catch the strong correlations induced by the hard core of the Argonne interaction, while NNLO$_{\rm sat}$, that is a relatively soft potential, may be more amenable to this approximate mapping onto LDA.
Further investigations of this hypothesis will require seeking an appropriate a quantitative measure of the hardness of these interactions~\cite{rios2017}. Energy densities, although they are not observables, may be useful to link ab initio and DFT~\cite{atkinson2020}. For example, they may help clarifying to what extent ab initio calculations meet the hypothesis that underlie LDA.

We find that the quality of predictions obtained in the present exploratory work is promising, in particular for the NNLO$_{\rm sat}$ interactions. Therefore, we will aim at extending our work in several directions. 
First of all, future studies will focus on constraining the gradient terms systematically on ab initio, with the help of simulations of model systems such as neutron/proton drops \cite{Gandolfi2011} or perturbed nuclear matter \cite{Buraczynski2017}.
Also, some of the surface terms of empirical EDFs give rise to the so-called effective mass, which is known to be crucial to the description of the nuclear spectroscopy.
A second parallel line of development will aim at exploiting more refined statistical tools.
Bayesian statistics \cite{vonToussaint2011} or machine learning techniques \cite{MEHTA20191} can play a determinant role in the calibration and assessment of the EDFs. Providing error bars on the EDF predictions would be a very important step forward.
Lastly, we plan to apply these EDFs on a wider range of physical problems, including time-dependent DFT. 

%% file: acknowledgement.tex
\section{Acknowledgements}
We thank Arianna Carbone for providing the NNLO$_{\rm sat}$ equations of state.  
SCGF and AFDMC calculations were performed using HPC resources at the DiRAC DiAL system at the University of Leicester, United Kingdom (BIS National E-infrastructure Capital Grant No. ST/K000373/1 and STFC Grant No. ST/K0003259/1).
Monte Carlo numerical calculations were made possible also through a CINECA-INFN agreement, providing access to resources on MARCONI at CINECA.
Funding from the European Union's Horizon 2020 research and innovation programme under grant agreement No 654002 is acknowledged. This work was supported in part by the U.S.\ Department of Energy (DOE), Office of Science, Office of Nuclear Physics, by the Argonne LDRD program, and by the NUCLEI project under contract number DE-AC02-06CH11357. 
A.L. was also supported by DOE Early Career Research Program awards.

%% file: appendix.tex
\section{LDA mean field potential}
\label{app: derivation mean field}
We derive the mean field potential $U_q(\varR)$ for the LDA EDF Sec. \ref{sec: edf lda}. By definition:
\begin{equation}
    U_q(\varR) = \fdv{E_{bulk}}{\rho_q(\varR)} =
    \pdv{\mathcal{E}_{bulk}}{\rho_q}  =
    \pdv{\mathcal{E}_{bulk}}{\rho} +
    \pdv{\beta}{\rho_q} \pdv{\mathcal{E}_{bulk}}{\beta}
\end{equation}
with q=n, p. 
Using $\rho=\rho_n+\rho_p$ and $\beta=\frac{\rho_n-\rho_p}{\rho}$, the chain rule leads to the following contributions:
\begin{align}
      & \pdv{\beta}{\rho_q} = \frac{1}{\rho} \,\pdv{}{\rho_q} \left( \rho_n - \rho_p \right) + \\ 
     & \left( \rho_n - \rho_p \right)   \left( -\frac{1}{\rho^2}\right) \pdv{\rho}{\rho_q} = 
     \frac{\tau_z-\beta}{\rho}   \nonumber   \\
     & \pdv{\mathcal{E}_{loc}}{\rho} = \sum_\gamma \left( \gamma+1 \right) c_\gamma(\beta) \rho^\gamma \\
    & \pdv{\mathcal{E}_{loc}}{\beta} =
    \sum_\gamma \pdv{c_\gamma(\beta)}{\beta} \rho^\gamma =
    2 \beta \sum_\gamma c_{\gamma,1} \rho^{\gamma+1}
\end{align}
where $\tau_z=+1$ for neutrons and $\tau_z=-1$ for protons.
Therefore
\begin{align*}
    U_q = \sum_\gamma (\gamma+1) c_\gamma(\beta) \rho^\gamma + (\tau_z-\beta) \, 2\beta \sum_\gamma c_{\gamma,1} \rho^\gamma =
    \nonumber \\
    \sum_\gamma \Big[ 
    \left(  \gamma+1\right) c_\gamma(\beta) \rho^\gamma + 2 \beta \left( \tau_z -\beta \right) c_{\gamma,1}
    \Big] \rho^\gamma
\end{align*}
and finally
\begin{equation*}
    U_q(\varR) = \sum_\gamma \Big[
    \left(  \gamma+1 \right) c_{\gamma,0}  + 2 \beta \left( \tau_z -\beta \right) c_{\gamma,1} +
    (\gamma+1) c_{\gamma,1} \beta^2 
    \Big] \rho^\gamma,
\end{equation*}
which proves Eq. \eqref{eq: mean field LDA}.

\section{GA mean field potential}
\label{app: mean field ga}
The mean field $U_q^{surf}$ Eq. \eqref{eq: mean field surf} is derived. By definition, $U_q^{surf}(\varR) = \fdv{E_{surf}}{\rho_q}$, where $E_{surf}$ is conveniently written as the volume integral of the density:
\begin{align*}
    \mathcal{E}_{surf} = & - \sum_{t=0,1} C_t^\Delta \abs{\nabla \rho_t}^2 \\
    & - \frac{W_0}{2} \left( \rho \nabla\cdot\mathbf{J} + \rho_n \nabla\cdot\mathbf{J}_n + \rho_p \nabla\cdot\mathbf{J}_p \right) 
\end{align*}
Then
\begin{equation}
    \fdv{E_{surf} }{\rho_q} (r) = \pdv{\mathcal{E}_{surf}}{\rho_q}
    - \nabla \cdot \left( \pdv{\mathcal{E}_{surf}}{ \left(\nabla\rho_q\right)} \right).
\end{equation}
The first contribution is due to the spin-orbit part and is equal to
\begin{equation}
\label{eq: mf part 1}
    \pdv{\mathcal{E}_{surf}}{\rho_q} = - \frac{W_0}{2} \left( \nabla \cdot \mathbf{J} + \nabla \cdot \mathbf{J}_q \right).
\end{equation}
The second one is due to the gradient terms of the EDF. To compute it, we first insert $\rho_0=\rho_n+\rho_p$ and \hbox{$\rho_1=\rho_n-\rho_p$} into the energy density:
\begin{align*}
    \mathcal{E}_{surf} & = 
    -\left( C_0^\Delta + C_1^\Delta \right) \left( \abs{\nabla \rho_n}^2 + \abs{\nabla \rho_p}^2 \right) + \\ & + 2 \left( C_1^\Delta - C_0^\Delta \right) \nabla\rho_n \cdot \nabla\rho_p,
\end{align*}
and then take the derivative:
\begin{align}
\label{eq: mf part 2}
    & - \nabla \cdot \left( \pdv{\mathcal{E}_{surf}}{ \left(\nabla\rho_q\right)} \right) = \\ &
    2 \left( C_0^\Delta + C_1^\Delta \right) \Delta \rho_q 
    - 2 \left( C_1^\Delta - C_0^\Delta \right) \Delta\rho_{\bar{q}} =  \nonumber \\ &
    2 C_0^\Delta \left( \Delta \rho_q + \Delta\rho_{\bar{q}} \right) +
    2 C_1^\Delta \left( \Delta \rho_q - \Delta\rho_{\bar{q}} \right) =
    \nonumber \\
    & 2 C_0^\Delta \Delta\rho_0 + 2 C_1^\Delta \Delta\rho_1 \tau_z \nonumber
\end{align}
where $\bar{q}=p$ if $q=n$ and viceversa. Summing Eqs. \eqref{eq: mf part 1} and \eqref{eq: mf part 2} concludes the derivation. 

\section{Rearrangement energy}
\label{app: derivation rearrangement}
In nuclear DFT, the total energy can be computed in two independent ways:
\begin{itemize}
    \item As the space integral of the EDF evaluated on the ground state densities that one obtains by solving the mean field equations:
    \begin{equation}
        E = \int d\varR \, \mathcal{E}(\varR)
    \end{equation}
    \item With the Hartree-Fock (HF) formula for a density-dependent Hamiltonian \cite{ring}:
    \begin{equation}
        E = \frac{1}{2} \left( T+ \sum_k \epsilon_k \right) + E_{rea}
    \end{equation}
\end{itemize}
The extra term $E_{rea}$ is called rearrangement energy.
The equality of the two expressions for the binding energy is often used as a non-trivial test of
the correctness of the implementation of a DFT or HF code.

 Here, the rearrangement energy for the LDA EDF of Sec. \ref{sec: edf lda} is derived. The following practical definition is employed:
\begin{equation}
    E_{rea} = \int d\varR\, \mathcal{E}_{bulk}(\varR) - \frac{1}{2} \sum_q \int d\varR U_q(\varR) \rho_q(\varR)
\end{equation}
with the mean field $U_q(\varR)$ \ref{eq: mean field LDA}. Then:
\begin{align*}
    & \sum_q U_q \rho_q = U_n \rho_n + U_p \rho_p = \\
    & \left[
    U_n (1+\beta) + U_p (1-\beta) \right] \frac{\rho}{2} = 
    \left[ \left( U_n + U_p \right) +
    \left( U_n - U_p \right) \beta
    \right] \frac{\rho}{2} \nonumber
\end{align*}
We calculate $U_n + U_p$ and $U_n - U_p$ separately:
\begin{align*}
     U_n+U_p & = 2 \sum_\gamma \left[ 
    \left( \gamma+1 \right) c_{\gamma,0} - 2 \beta^2 c_{\gamma,1} + \left( \gamma+1 \right) \beta^2 c_{\gamma,1}
    \right] \rho^\gamma =  \\
    & = 2 \sum_\gamma \left[ \left( \gamma+1 \right) c_{\gamma,0} + \left( \gamma-1 \right) \beta^2 c_{\gamma,1} \right] \rho^\gamma \nonumber 
\end{align*}
and 
\begin{equation*}
     U_n -U_p = 4 \beta  \sum_\gamma c_{\gamma,1} \rho^\gamma
\end{equation*}
with $\tau_z=1$ for neutrons and $\tau_z=-1$ for protons.
Then:
\begin{equation*}
    \sum_q U_q \rho_q = \frac{\rho}{2} \, 2 \sum_\gamma
    \left[  \left( \gamma+1 \right) c_{\gamma,0} + \left( \gamma+1 \right) \beta^2 c_{\gamma,1}
    \right] \rho^{\gamma+1}
\end{equation*}
Plugging into the definitions of $E_{rea}$:
\begin{align*}
    E_{rea} & = \int d\varR \, \sum_\gamma \,
    \Big[
    \left( c_{\gamma,0} + \beta^2 c_{\gamma,1} 
    \right) \rho^{\gamma+1} + \\
    & -
    \left( \frac{1 + \gamma}{2} \right) \left( c_{\gamma,0} + \beta^2 c_{\gamma,1}
    \right) \rho^{\gamma+1} 
    \Big]
\end{align*}
and finally:
\begin{equation}
    E_{rea} = \int d\varR \sum_\gamma \left(
    \frac{1-\gamma}{2}
    \right) \left( c_{\gamma,0} + \beta^2 c_{\gamma,1}
    \right) \rho^{\gamma+1}
\end{equation}